\documentclass[a4paper,11pt,oneside]{article}
\usepackage[pdftex]{graphicx,color}      
\usepackage{verbatim}
\usepackage[T1]{fontenc}
\usepackage{amsfonts}    
\usepackage{textcomp}
\usepackage[dvips]{epsfig}
\usepackage[para, symbol]{footmisc}
\usepackage{threeparttable}
\usepackage[font=small]{caption}
\usepackage{setspace}

\begin{document}

\begin{center}
\Large\textbf{How orthogonalities set Kochen-Specker sets}
\end{center}

\begin{center}
\large Kate Blanchfield\footnote{kate@fysik.su.se}
\end{center}

\begin{center}
\large\textit{Stockholms universitet, Fysikum, 106 91 Stockholm, Sweden}
\end{center}

\begin{abstract}
We look at generalisations of sets of vectors proving the Kochen-Specker theorem in 3 and 4 dimensions. It has been shown that two such sets, although unitarily inequivalent, are part of a larger 3-parameter family of vectors that share the same orthogonality graph. We find that these sets are unusual, in that the vectors in all other Kochen-Specker sets investigated here are fully determined by orthogonality conditions and thus admit no free parameters.
\end{abstract}

\paragraph\noindent The Kochen-Specker (KS) theorem \cite{Kochen Specker} is a statement about the fundamental nature of quantum mechanics: no non-contextual hidden variable theory can reliably reproduce quantum mechanical predictions. The KS theorem states that in a Hilbert space with dimension $n \geq 3$ it is impossible to assign definite values $v \in\left\{ 0,1 \right\}$ to $m$ projection operators $P_m$ so that for every \textit{commuting} set of projection operators satisfying $\sum P_m = \mathbb{I}$, the corresponding definite values obey $\sum v \left( P_m \right) = 1$.
\newline\indent The KS theorem is often visualised by means of a set of "uncolourable" vectors, where projection operators are represented by their constituent rays in Hilbert space and rays can be simplified to vectors as only directions are relevant here. The vectors are coloured according to their pre-assigned values and the summation conditions above translate into colouring rules. This allows us to construct orthogonality graphs that, when uncolourable, provide a contradiction within non-contextual hidden variable models and thus a proof of the KS theorem.

\paragraph\noindent Sets of vectors exemplifying the KS theorem have been discovered by numerous authors, with an aim to reduce the number of vectors required to reach a contradiction. Most notably, in a 3-dimensional Hilbert space, the original KS proof of 117 vectors has been replicated with sets using between 31 and 33 vectors \cite{Peres, Penrose, Bub, Conway Kochen} and in a 4-dimensional Hilbert space, sets of between 24 and the lower limit of 18 vectors have been realised \cite{Peres, Kernaghan, Cabello}. Additionally, an extensive computer search has recently revealed over one thousand of such KS sets in $\mathbb{R}^4$ \cite{Pavicic2}.

\paragraph\noindent In a publication by E. Gould and P. K. Aravind \cite{Gould Aravind}, two of these sets in 3 dimensions have been shown to share the same orthogonality graph. Specifically, the sets of 33 vectors found by Penrose \cite{Penrose} and Peres \cite{Peres} have identical orthogonality conditions but are not unitarily equivalent. It is easy to see there can be no unitary operation to reconcile the sets as the Penrose set has vector entries from $\mathbb{C}^3$ and cannot be confined to a real subspace while the Peres set is entirely real. Furthermore, both sets are shown in \cite{Gould Aravind} to be special cases of a more general 3-parameter family of 33 vectors.
\newline\indent This observation can be applied to other KS sets, leading to the question: how common are such parameterised generalisations among KS sets? In effect, we are asking to what extent the orthogonality conditions between vectors in KS sets determine the vectors themselves.

\paragraph\noindent KS sets were reconstructed with their most general vectors through a straight-forward procedure using their orthogonality graphs. Global phase freedom was exploited to rotate one orthogonal basis (triad in $\mathcal{H}^3$ or tetrad in $\mathcal{H}^4$) into the standard basis, and all subsequent vectors were formulated solely from given orthogonality constraints. The parameters inevitably introduced into the vectors initially, became constrained and eventually solvable through later orthogonality conditions.
\newline\newline\indent The KS sets studied are listed in table~\ref{tab:a} with additional properties of each set. These properties refer only to the vectors given in the original proofs, although in some of these cases additional vectors are implied from rotations. The increase in the number of vectors and subsequently in the number of orthogonalities and bases in such extended sets has been calculated by Larsson \cite{Larsson}. The result of this study, namely the number of parameters present in the generalisation of original KS set vectors, is shown in the final column of table~\ref{tab:a}.

\paragraph\noindent
\begin{center}
\small
\begin{threeparttable}
\begin{tabular}{l c c c c c}
\hline
KS set & n\tnote{a} & Vectors & $\perp$\tnote{b} & Bases & Parameters\\
\hline\hline
Peres \cite{Peres} / Penrose \cite{Penrose} & 3 & 33\tnote{c} & 72 & 16 & 3\\
Schütte \cite{Bub} & 3 & 33\tnote{c} & 76 & 20 & 0\\
Conway, Kochen \cite{Conway Kochen} & 3 & 31\tnote{c} & 70 & 17 & 0\\
Peres \cite{Peres} & 4 & 24 & 144 & 24 & 0\\
Kernaghan \cite{Kernaghan} & 4 & 20\tnote{c} & 73 & 11 & 0\\
Pavi\v{c}i\'{c} \textit{et al.} \cite{Pavicic1} & 4 & 20\tnote{c} & 75 & 11 & 0\\
Cabello \textit{et al.} \cite{Cabello} & 4 & 18\tnote{c} & 54 & 9 & 0\\
Pavi\v{c}i\'{c} \textit{et al.} \cite{Pavicic1} & 4 & 24 & 154 & 13 & 0\\
\hline
\end{tabular}
\caption{Parameters reported in this article shown with other properties of the KS sets.}
\label{tab:a}
\begin{tablenotes}
\item[a] Dimension
\item[b] Number of orthogonality conditions
\item[c] Critical set (for a description of critical sets see, for instance, \cite{Aravind})
\end{tablenotes}
\end{threeparttable}
\end{center}

\normalsize
\paragraph\noindent
All vectors are fully determined by their orthogonality relations for the KS sets reported here, thus we find no free parameters. There are no obvious trends in the data shown in table~\ref{tab:a} to account for this surprising lack of parameters. This finding adds a curious apparent uniqueness to the Peres and Penrose sets in $\mathcal{H}^3$ and thus strengthens the appeal of the isomorphism found in \cite{Gould Aravind}.
\newline\indent This study does not claim to have performed an exhaustive search, having examined only seven of the many KS sets available, but it highlights a new issue, namely the variation in the freedom of vector choice for KS sets.

\setstretch{0.9}


\begin{thebibliography}{99}
\bibitem{Kochen Specker} S. Kochen and E. P. Specker, "The Problem of Hidden Variables in Quantum Mechanics" \textit{J. Math. Mech.} \textbf{17} 59-87 (1967)
\bibitem{Peres} A. Peres, "Two simple proofs of the Kochen-Specker theorem" \textit{J. Phys. A} \textbf{24} 175-178 (1991)
\bibitem{Penrose} R. Penrose, "On Bell non-locality without probablities: some curious geometry", in \textit{Quantum Reflections} edited by J. Ellis and D. Amati, Cambridge University Press (2000)
\bibitem{Bub} J. Bub, "Schütte's Tautology and the Kochen-Specker Theorem" \textit{Foundations of Physics} \textbf{26} 787-806 (1996)
\bibitem{Conway Kochen} Reported in A. Peres, \textit{Quantum Theory: Concepts and Methods}, Kluwer Academic Publishers, page 114 (1998)
\bibitem{Kernaghan} M. Kernaghan, "Bell-Kochen-Specker theorem for 20 vectors" \textit{J. Phys. A: Math. Gen} \textbf{27} L829-L830 (1994)
\bibitem{Cabello} A. Cabello, J. Estebaranz and G. García-Alcaine, "Bell-Kochen-Specker theorem: A proof with 18 vectors" \textit{Phys. Lett. A} \textbf{212} 183 (1996)
\bibitem{Pavicic2} M. Pavi\v{c}i\'{c}, N. D. Megill and J.-P. Merlet, \textit{arXiv:quant-ph/0910.1311v4} (2010)
\bibitem{Pavicic1} M. Pavi\v{c}i\'{c}, J.-P. Merlet, B. McKay and N. D. Megill, "Kochen-Specker vectors" \textit{J. Phys. A: Math. Gen} \textbf{38} 1577-1592 (2004)
\bibitem{Gould Aravind} E. Gould and P. K. Aravind, "Isomorphism between the Peres and Penrose proofs of the BKS theorem in three dimensions" \textit{arXiv:quant-ph/0909.4502v2}
\bibitem{Larsson} J.-Å. Larsson, "A Kochen-Specker inequality" \textit{Europhys. Lett.} \textbf{58} 799-805-87 (2002)
\bibitem{Aravind} P. K. Aravind, "How Reye's configuration helps in proving the Bell-Kochen-Specker theorem: a curious geometrical tale" \textit{Found. Phys. Lett.} \textbf{13} 499-519 (2000)
\end{thebibliography}
\end{document}